\documentstyle{elsart}
\begin{document}

\begin{frontmatter}
\hfill preprint INP 1816/PH

\title{Subtracted Dispersion Relations for In-Medium Meson Correlators in QCD
Sum Rules\thanksref{grants}}
\thanks[grants]{Research supported by
        the Polish State Committee for
        Scientific Research, grant 2P03B-080-12.}
\thanks[emails]{E-mail: florkows@solaris.ifj.edu.pl,
 broniows@solaris.ifj.edu.pl}
\author{Wojciech Florkowski} and
\author{Wojciech Broniowski}
\address{H. Niewodnicza\'nski Institute of Nuclear Physics, ul.
         Radzikowskiego 152, PL-31342 Krak\'ow, Poland}
\begin{abstract}
We analyze subtracted dispersion relations for meson
correlators at finite baryon density
and temperature. Such relations are needed for QCD sum rules. We point out
the importance of scattering terms, as well as finite, well-defined
subtraction constants. Both are necessary for consistency,
in particular for the
equality of the longitudinal and transverse correlators in the limit of the
vanishing three-momentum of mesons relative to the medium. We present
detailed calculations in various mesonic channels for the case of
the Fermi gas of nucleons.
\end{abstract}

\begin{keyword}
in-medium meson correlators, vector mesons, QCD sum rules in nuclear matter,
dispersion relations
\end{keyword}
\end{frontmatter}
\vspace{-7mm} PACS: 21.65.+f, 11.55.Hx

\section{Introduction}

\label{sec:Intro}There has been a continued interest in the studies of
in-medium properties of hadrons. As the QCD ground state is largely modified
at finite temperature and/or baryon density, one expects that masses,
widths, coupling constants, and other characteristics of hadrons change
considerably \cite{heidelberg,hadrons}. In particular, the properties of
light vector mesons are of the special interest, since their modification
may influence dilepton spectra measured in relativistic heavy-ion
collisions. In fact, the observed excess of dilepton pairs in such
experiments \cite{ceres,helios} is commonly explained by the assumption that
the masses of the vector mesons decrease or their widths become larger, or
both \cite{li,cassing,rapp,brat}.

Different theoretical methods have been used
to analyze medium effects on hadron properties. Let us mention
effective hadronic models \cite{herrmann,klingl}, low-density theorems
\cite{eletsky,friman}, quark models \cite{hatkun}, and QCD sum rules
\cite{BS,FHL,AHZ,HKL,Jin,Lissia,Loewe,Mallik,HL,HLS,LPM,LM}.
The results of this paper are relevant to this last method. In QCD sum
rules one needs the in-medium dispersion relation for the meson
correlator. Furtheremore, this dispersion relation is subtracted in
order to improve the reliability of the sum rule method. We point out
that the existent calculations
\cite{AHZ,HKL,Jin,Mallik,HL,HLS,LPM} have not been very careful in this
issue. They include the contributions form the so-called scattering
term in the spectral density, but overlook the fact that in addition
there is a well-defined, finite subtraction constant in the dispersion
relation that has to be kept. This subtraction constant, which depends
on the channel, is crucial for consistency of the approach.
For instance, only when it
is kept then the transverse and longitudinal correlators in vector and
tensor channels are equal in the case of the vanishing three-momentum
in the medium rest frame (${\bf q}=0$). Since there seems to be quite
a deal of confusion in the literature, we are are trying to be very
explicit in this paper, showing the calculation of meson correlators in
a Fermi gas of nucleons at ${\bf q}=0$.

So far, the majority of the applications of the QCD-sum-rule method has
dealt with the vector mesons. In this paper, having in mind possible future
applications to other mesons, we collect the formulas for different channels
as well.

\section{Dispersion relations and QCD sum rules}
\label{sect:Disp}

QCD sum rules \cite{SVZ,RRY} for meson correlators at finite
temperature \cite{BS,FHL,AHZ,HKL,Jin,Lissia,Loewe,Mallik} and finite
density \cite{HL,HLS,LPM,LM} have provided useful information on
medium modifications of mesonic spectra. The basic quantity in these
studies is the retarded in-medium correlator of two meson currents,
$J$ and $J^{\prime }$, defined as
\begin{equation}
\Pi (\omega ,{\bf q})=i\int d^4xe^{i\omega t-i{\bf q}\cdot {\bf x}}\theta
(t)\left\langle \left\langle [J(x),J^{\prime }(0)]\right\rangle
\right\rangle ,  \label{cordef}
\end{equation}
where $\left\langle \left\langle ...\right\rangle \right\rangle $ denotes
the Gibbs average. Since the medium breaks the Lorentz invariance, the
correlator depends separately on the energy variable, $\omega $, and the
momentum, ${\bf q}$, measured in the rest frame of the medium. The
correlator (\ref{cordef}) satisfies the usual fixed-${\bf q}$ dispersion
relation \cite{HKL,Lissia},
\begin{equation}
\Pi (\omega ,{\bf q})=\frac 1\pi \int_0^\infty
d\nu ^2\frac{\rho (\nu ,{\bf q})}{\nu
^2-\omega ^2-i\varepsilon \hbox{sgn}(\omega )},  \label{disprel}
\end{equation}
with $\rho (\nu ,{\bf q})={\rm Im}\Pi (\nu ,{\bf q})$ denoting the
spectral density. The RHS is not well-defined and the relation
requires subtractions in order for the integral to converge. In QCD
sum rules one applies the Borel transform, $\hat L_{M_B}$
\cite{SVZ,RRY}, to both sides of Eq. (\ref{disprel}), where $M_B$
denotes the Borel mass parameter. The procedure is carried as follows:
first the dispersion relation (\ref{disprel}) is continued to the
Euclidean space by replacing $\omega $ with $i\Omega $, and then the
Borel transform is carried in the variable $\Omega ^2$. As a result
the RHS of Eq.  (\ref{disprel}) becomes $\frac 1{\pi M_B^2}\int d\nu
^2e^{-\nu ^2/M_B^2}\rho (\nu ,{\bf q})$, which is well-defined. Thus,
in fact, one does not need to perform subtractions in
Eq. (\ref{disprel}), since the Borel transform makes the spectral
integral convergent. However, one can still perform the subtractions
and in fact it is advantageous to do so in order to obtain more
reliable sum
rules for vector mesons both in the vacuum \cite{SVZ,RRY} and in
medium \cite{LM}. The reason is that subtractions reduce the
contribution from the high-lying resonances and continuum states in
the spectral integral, as can be seen from the formula given
below.

The existent calculations of in-medium {\em vector mesons}
\cite{Jin,HL,HLS} have implicitly
made use of the following subtracted dispersion
relation, which follows from Eq. (\ref{disprel}) when the subtraction
is made at the point $\omega =|{\bf q}|$
\begin{equation}
\frac{\Pi (\omega ,{\bf q})-\Pi (|{\bf q}|,{\bf q})}{\omega ^2-{\bf q}^2}%
=\frac 1\pi \int_0^\infty
d\nu ^2\frac{\rho (\nu ,{\bf q})}{(\nu ^2-{\bf q}^2)(\nu
^2-\omega ^2)}.  \label{subtr}
\end{equation}
(We disregard the infinitesimal imaginary $i \varepsilon$
terms here, since they are
irrelevant in QCD-sum-rule applications.) Compared to the study in the
vacuum, there are two important differences. The first one is the appearance
of the $\Pi (|{\bf q}|,{\bf q})$ term, which is zero in the vector meson
correlator in the vacuum, but in general {\em does not vanish in medium}
(for instance, this is the case of the transverse vector meson correlator
described in Section \ref{sect:piqq}). The second difference is related to the
analytic structure of the spectral density $\rho (\nu ,{\bf q})$ in nuclear
medium. In addition to the singularities in the time-like region ($\nu >|%
{\bf q}|$), the presence of medium induces singularities in the space-like
region, ($\nu <|{\bf q}|$) \cite{chin}. For instance, the particle-hole
excitations of the nucleon Fermi sea at vanishing temperature lead to a cut
reaching in the region $0\leq $ $\nu \leq |{\bf q}|k_F/E_F$, where $k_F$ and 
$E_F$ are the Fermi momentum and energy. The spectral density connected
with this cut is denoted by $\rho _{sc}(\nu ,{\bf q}%
) $. This phenomenon is called the {\em scattering correction}, or the {\em %
Landau damping} contribution, and its relevance for QCD sum rules was first
pointed out by Bochkarev and Shaposhnikov \cite{BS} for the case of $\rho $
mesons propagating in a pion gas. It has been subsequently used in
Refs.~\cite{FHL,AHZ,HKL,Jin,Mallik,HL,HLS,LPM}.

The contribution of the
scattering term is subtle in the limit of ${\bf q}\rightarrow 0$, which
describes the mesonic excitations at rest with respect to the nuclear
medium. The point is that in this limit one finds
\begin{equation}
\lim_{{\bf q}\rightarrow 0}\frac{\rho _{sc}(\nu ,{\bf q})}{\nu ^2-{\bf q}^2}%
=R\delta (\nu ^2),  \label{delta}
\end{equation}
where $R$ is a constant depending on the channel. The following sections
provide a detailed analysis of this point.
Thus, in the limit of ${\bf q}\rightarrow 0$ the following Borel-improved
sum rule results from the subtracted dispersion relation (\ref{subtr})
\begin{equation}
\hspace{-1.0cm}\hat L_M\left( \frac{\Pi (i\Omega ,0)}{-\Omega ^2}\right)
=-\frac{\lim_{{\bf q}\rightarrow 0}\Pi (|{\bf q}|,{\bf q})}{M_B^2}\!+\!
\frac R{\pi M_B^2}\!+\!\frac
1{\pi M_B^2}\int_0^\infty
d\nu ^2e^{-\nu ^2/M_B^2}\frac{\bar \rho (\nu ,0)}{\nu ^2},
\label{borel0}
\end{equation}
where we have explicitly separated out the scattering term, and introduced $%
\bar \rho (\nu ,0)\equiv \rho (\nu ,{\bf q})-\rho _{sc}(\nu ,{\bf q})$. Note
that if (and only if) $\Pi (|{\bf q}|,{\bf q})$ vanishes, then Eq. (\ref
{subtr}) may be viewed as the unsubtracted dispersion relation for the
function $\tilde \Pi (\omega ,{\bf q})=\Pi (\omega ,{\bf q})/(\omega
^2-{\bf q}^2)$. This is the case of the vector correlator in the vacuum, or
of the longitudinal vector correlator in medium. But this is not the case in
other channels, in particular, $\Pi (|{\bf q}|,{\bf q})$ does not vanish for
the transverse vector channel in medium. In the following sections the issue
will be discussed in detail on an example where the medium consists of the
Fermi gas of nucleons, and the meson correlators are evaluated at the 1p-1h
level. This case is relevant, since existent QCD-sum-rule calculations at
finite density use this approximation to model the scattering term in the
phenomenological spectral function. A similar analysis can be carried for
more complicated many-body treatment, and for other contributions to the
correlators, such as the meson loops.

\section{Meson correlators in the Fermi gas of nucleons}

\label{sect:FermiGas}From now on our medium is the nuclear matter described
by the Fermi gas of nucleons. Using the imaginary-time formalism \cite
{FW,KAP} we evaluate the in-medium part of the meson correlator at the 1p-1h
level. A straightforward calculation leads to the formula
\begin{eqnarray}
\!\!\!\!\!\Pi ^{ab}(q) &=& - \int {\frac{d^3k}{2E_k(2\pi )^3}}\left\{ {\frac
1{e^{\beta (k\cdot u-\mu )}+1}}\left[ {\frac{{\cal T}_{ab}(k,q)}{q^2+2k\cdot
q}}+{\frac{{\cal T}_{ab}(k-q,q)}{q^2-2k\cdot q}}\right] \right.  \nonumber \\
&&\ \left. +{\frac 1{e^{\beta (k\cdot u+\mu )}+1}}\left[ {\frac{{\cal T}%
_{ab}(-k-q,q)}{q^2+2k\cdot q}}+{\frac{{\cal T}_{ab}(-k,q)}{q^2-2k\cdot q}}%
\right] \right\} _{k^0=E_k}.  \label{pidef}
\end{eqnarray}
Here $q$ is the
meson four-momentum,
$k$ is the internal four-momentum,  $k_0=E_k=\sqrt{M^2+{\bf k}^2}$,
$M$ is the nucleon mass, $u^\mu $ is the
four-velocity of the medium, $\beta $ is the inverse temperature,
$\mu $ is the
baryon chemical potential, and $a$ and $b$ label the channel. The Dirac
trace factor is given by
\begin{equation}
{\cal T}_{ab}(k,q)=\hbox{tr}\left[ (\FMSlash{k}+M)\Gamma _a(\FMSlash{k}+%
\FMSlash{q}+M)\Gamma _b\right] ,  \label{tfdef}
\end{equation}
where $\Gamma _a$ and $\Gamma _b$ denote the appropriate vertices ({\em cf.}
Table \ref{table1}). Note that in the rest frame of the medium $q=(\nu ,{\bf q})$,
where $\nu $ is a purely imaginary bosonic Matsubara frequency, $\nu =2\pi
in/\beta $, hence the denominators in (\ref{pidef}) are well defined. At a
later stage of the calculation we shall perform the standard analytic
continuation in order to obtain the imaginary part of the retarded
correlator. This is achieved by the substitution $\nu =2\pi in/\beta
\rightarrow \nu _R+i\varepsilon $, where $\nu _R$ is real.
\begin{table}[tb]
\begin{tabular}{|p{6.25cm}|p{1.25cm}|p{1.25cm}|p{1.25cm}|p{1.5cm}|}
\hline
\multicolumn{1}{|c|}{$i^{th}$ channel} & \multicolumn{1}{|c|}{index} &
\multicolumn{1}{|c|}{$\Gamma_a$} & \multicolumn{1}{|c|}{$\Gamma_b$} &
\multicolumn{1}{|c|}{${\cal P}^{ab}_{(i)}$} \\ \hline\hline
scalar & $S$ & 1 & 1 & 1 \\ \hline\hline
pseudoscalar & $P$ & $i \gamma_5$ & $i \gamma_5 $ & 1 \\ \hline\hline
longitudinal vector & $V_L$ & $\gamma_{\mu}$ & $\gamma_{\nu} $ & $L^{\mu
\nu} $ \\ \hline
transverse vector & $V_T$ & $\gamma_{\mu}$ & $\gamma_{\nu}$ & $T^{\mu \nu}$
\\ \hline\hline
$q$-longitudinal axial vector & $A_{qL}$ & $\gamma_{\mu}\gamma_5$ & $%
\gamma_{\nu}\gamma_5 $ & $Q^{\mu \nu}$ \\ \hline
longitudinal axial vector & $A_L$ & $\gamma_{\mu}\gamma_5$ & $%
\gamma_{\nu}\gamma_5 $ & $L^{\mu \nu}$ \\ \hline
transverse axial vector & $A_T$ & $\gamma_{\mu\gamma_5}$ & $%
\gamma_{\nu}\gamma_5 $ & $T^{\mu \nu}$ \\ \hline\hline
longitudinal tensor (parity --) & $T_{L^-}$ & $\sigma_{\mu \nu}$ & $%
\sigma_{\alpha \beta}$ & $L_{(-)}^{\mu \nu; \alpha \beta}$ \\ \hline
transverse tensor (parity --) & $T_{T^-}$ & $\sigma_{\mu \nu}$ & $%
\sigma_{\alpha \beta}$ & $T_{(-)}^{\mu \nu; \alpha \beta}$ \\ \hline
longitudinal tensor (parity +) & $T_{L^+}$ & $\sigma_{\mu \nu}$ & $%
\sigma_{\alpha \beta}$ & $L_{(+)}^{\mu \nu; \alpha \beta}$ \\ \hline
transverse tensor (parity +) & $T_{T^+}$ & $\sigma_{\mu \nu}$ & $%
\sigma_{\alpha \beta}$ & $T_{(+)}^{\mu \nu; \alpha \beta}$ \\ \hline\hline
vector-tensor mixing (longitudinal) & $VT_L$ & $\gamma_{\mu}$ & $%
\sigma_{\alpha \beta}$ & $L^{\mu ; \alpha \beta}$ \\ \hline
vector-tensor mixing (transverse) & $VT_T$ & $\gamma_{\mu}$ & $%
\sigma_{\alpha \beta}$ & $T^{\mu ; \alpha \beta}$ \\ \hline
\end{tabular}
\caption{List of channels considereded, with the corresponding
vertices $\Gamma _a$,$\Gamma _b$, and projection tensors
${\cal P}_{(i)}^{ab}$.}
\label{table1}
\end{table}

Except for the scalar and pseudoscalar channels we need to project
expression (\ref{pidef}) on the appropriate quantum numbers using projection
tensors ${\cal P}_{(i)}^{ab}$:
\begin{equation}
\Pi ^{ab}= \sum_i \Pi _{(i)}{\cal P}_{(i)}^{ab},\hspace{1cm}\Pi _{(i)}=\Pi _{ab}{%
\frac{{\cal P}_{(i)}^{ba}}{\hbox{dim}{\cal P}_{(i)}}},\hspace{1cm}\hbox{dim}%
{\cal P}_{(i)}=\sum_c{\cal P}_{(i)\,\,c}^{\,c}.
\end{equation}
The considered choices for $\Gamma _a$, $\Gamma _b$ and the corresponding $%
{\cal P}_{(i)}^{ab}$ are listed in Table~\ref{table1}.
We analyze all diagonal channels
and one non-diagonal case: the vector-tensor channel. The explicit
expressions for ${\cal P}_{(i)}^{ab}$ are given in the Appendix. The scalar
functions ${\cal P}_{(i)}$ depend on two Lorentz scalars: $q^2=\nu ^2-{\bf q}%
^2$, and $q\cdot u=\nu $.

In the considered cases the trace factor (\ref{tfdef}) satisfies the
following symmetry relations\footnote{%
This is not a universal feature and it does not hold for certain
non-diagonal channels, which we do not take into consideration in this paper
({\em e.g.} the scalar-vector channel).}:

\begin{eqnarray}
{\cal T}_{ab}(k-q,q)={\cal T}_{ab}(k,-q)\hspace{0.25cm} &,&\hspace{0.25cm}%
{\cal T}_{ab}(-k-q,q)={\cal T}_{ab}(k,q),  \nonumber \\
{\cal T}_{ab}(-k,q) &=&{\cal T}_{ab}(k,-q).
\end{eqnarray}
This feature allows us to rewrite Eq. (\ref{pidef}) in the following form
\begin{equation}
\Pi ^{(i)}=\int {\frac{d^3k}{2E_k(2\pi )^3}}f(E_k,\mu ,\beta ){\frac{{\cal N}%
_{(i)}}{q^4-4(E_k\nu -{\bf k}\cdot {\bf q})^2}},  \label{pidef1}
\end{equation}
where
\begin{equation}
f(E_k,\mu ,\beta )={\frac 1{e^{\beta (E_k-\mu )}+1}}+{\frac 1{e^{\beta
(E_k+\mu )}+1}}  \label{distr}
\end{equation}
and
\begin{eqnarray}
\hspace{2cm} {\cal N}_{(i)}&=&
\left\{ \right. 2k\cdot q\,\left[ {\cal T}_{ab}(k,q)-{\cal T}_{ab}(k,-q)\right]
\nonumber \\
\hspace{2cm} &-&  \left. q^2\left[ {\cal T}_{ab}(k,q)+{\cal T}_{ab}(k,-q)\right]
\right\} {\frac{{\cal P}_{(i)}^{ab}} {\hbox{dim} {\cal P}_{(i)}}}.
\label{N}
\end{eqnarray}
We can decompose
\begin{equation}
{\cal N}_{(i)}=A_{(i)}+B_{(i)}x+C_{(i)}x^2,  \label{ABC}
\end{equation}
where $x$ is the the cosine of the angle between the three-vectors ${\bf k}$
and ${\bf q}$, and the functions $A_{(i)},B_{(i)}$ and $C_{(i)}$ depend only
on $|{\bf k}|,\nu $ and $|{\bf q}|$ . {\em For simplicity of notation we
shall from now on denote} $|{\bf k}|$ {\em and} $|{\bf q}|$ {\em by} $k$ 
{\em and} $q$, {\em respectively}. The explicit form of the coefficients $%
A_{(i)},B_{(i)}$ and $C_{(i)}$ for the considered channels is given in
Table \ref{table2} .

The angular integration in (\ref{pidef1}) is elementary and yields 
\begin{eqnarray}
&&\int_{-1}^1dx{\frac{A_{(i)}+B_{(i)}x+C_{(i)}x^2}{(\nu ^2-q^2)^2-4(E_k\nu
-kqx)^2}}={\frac 1{16k^3q^3}}\left\{ -8C_{(i)}kq\vphantom{{A \over B}}\right.
\nonumber \\
&&\left. +{\cal L}_{(i)}^{+}\ln \left[ {\frac{2E_k\nu +\nu ^2-q^2-2kq}{%
2E_k\nu +\nu ^2-q^2+2kq}}\right] +{\cal L}_{(i)}^{-}\ln \left[ {\frac{%
2E_k\nu -\nu ^2+q^2+2kq}{2E_k\nu -\nu ^2+q^2-2kq}}\right] \right\} , 
\nonumber \\
&&  \label{xintegral}
\end{eqnarray}
with 
\begin{eqnarray}
&&\!\!\!\!\!\!\!\!\!\!\!\!\!{\cal L}_{(i)}^{\pm }={\frac{C_{(i)}[2E_k\nu \pm
\nu ^2\mp q^2]^2+2kq\left[ 2A_{(i)}kq+B_{(i)}(2E_k\nu \pm \nu ^2\mp
q^2)\right] }{q^2-\nu ^2}}.  \nonumber \\
&&  \label{lpm}
\end{eqnarray}
By substituting (\ref{xintegral}) in (\ref{pidef1}) we express $\Pi ^{(i)}$
as a single integral over $k$.

\begin{table}[bt]
\begin{tabular}{|p{1.5cm}|p{3.8cm}|p{3.3cm}|p{3.3cm}|}
\hline
\multicolumn{1}{|c|}{$(i)$} & \multicolumn{1}{|c|}{$A_{(i)}/16$} & 
\multicolumn{1}{|c|}{$B_{(i)}/16$} & \multicolumn{1}{|c|}{$C_{(i)}/16$} \\ 
\hline\hline
$S$ & $k^2\nu ^2+M^2q^2$ & $-2E_kk\nu q$ & $k^2q^2$ \\ \hline\hline
$P$ & $E_k^2\nu ^2$ & $-2E_kk\nu q$ & $k^2q^2$ \\ \hline\hline
$V_L$ & $(q^2-\nu^2)E_k^2$ & $0$ & $(\nu ^2-q^2)k^2$ \\ \hline
$V_T$ & $\half (\nu ^2-q^2)k^2-E_k^2\nu ^2$ & $2E_kk\nu q$ & $-\half k^2(\nu
^2+q^2)$ \\ \hline\hline
$A_{qL}$ & $M^2(\nu ^2-q^2)$ & $0$ & $0$ \\ \hline
$A_L$ & $(q^2-\nu^2)k^2$ & $0$ & $(\nu ^2-q^2)k^2$ \\ \hline
$A_T$ & $-\half k^2(\nu ^2+q^2)-M^2q^2$ & $2E_kk\nu q$ & $-\half k^2(\nu
^2+q^2)$ \\ \hline\hline
$T_{L^{-}}$ & $-2(M^2\nu ^2+k^2q^2)$ & $4E_kk\nu q$ & $-2k^2\nu ^2$ \\ \hline
$T_{T^{-}}$ & $(q^2-\nu^2)(k^2+2M^2)$ & $0$ & $(\nu ^2-q^2)k^2$ \\ \hline
$T_{L^{+}}$ & $2E_k^2q^2$ & $-4E_kk\nu q$ & $2k^2\nu ^2$ \\ \hline
$T_{T^{+}}$ & $(\nu ^2-q^2)k^2$ & $0$ & $(q^2-\nu^2)k^2$ \\ \hline\hline
$VT_{L}$ & $-i2^{-1/2}(\nu ^2-q^2)^{3/2}M$ & $0$ & $0$ \\ \hline
$VT_{T}$ & $-i2^{-1/2}(\nu ^2-q^2)^{3/2}M$ & $0$ & $0$ \\ \hline
\end{tabular}
\caption{Explicit forms of the coefficients $A_{(i)},B_{(i)}$ and $C_{(i)}$
defined by Eqs. (\ref{N}) and (\ref{ABC}), here $q^2={\bf q} \cdot {\bf q}$
and $k^2={\bf k} \cdot {\bf k}$.}
\label{table2}
\end{table}

\section{Subtraction constant $\Pi (q,q)$}
\label{sect:piqq}Equipped with explicit formulas from the preceding section
we may now evaluate the subtraction constant $\Pi (q,q)$. By inspection of
Table \ref{table2}, we find that the ratio 
${\cal N}_{(i)}(k,\nu =q,q)/(E_kq-{\bf k}%
\cdot {\bf q})^2$ is a number depending only on the channel. It vanishes in
vector and axial-vector longitudinal channels, and in transverse tensor
channels. In other cases $\Pi (q,q)$ is given as an integral over the
distribution function (\ref{distr}). For example, in the transverse vector
channel we have
\begin{equation}
\Pi _{V_T}(q,q)={\frac 1{\pi ^2}}\int {\frac{dk\,k^2}{E_k}}f(E_k,\mu ,\beta
).  \label{piqqvt}
\end{equation}
At zero temperature the integral over the distribution function can be done
analytically with the result
\begin{eqnarray}
\hspace{1.5cm} \Pi ^{(i)}(q,q) &=&-{\frac 14}\int_0^{k_F}
{\frac{d^3k}{2E_k(2\pi )^3}}{\frac{%
{\cal N}_{(i)}(k,\nu =q,q)}{(E_kq-{\bf k}\cdot {\bf q})^2}}  \nonumber \\
\hspace{1.5cm} &\equiv&W^{(i)}\left[ {\frac{M^2}{\pi ^2}}\Phi (v_F)+
{\frac{k_F^3}{\pi ^2E_F}}%
\right] ,  \label{piqq}
\end{eqnarray}
where
\begin{equation}
\Phi (v_F)=v_F+\half \ln {\frac{1-v_F}{1+v_F}}  \label{phi}
\end{equation}
and $v_F=k_F/E_F=k_F/\sqrt{M^2+k_F^2}$ is the velocity of nucleons on the
Fermi surface. Note that in our calculation $\Pi ^{(i)}(q,q)$ does not
depend on $q$. The coefficients $W^{(i)}$ characterize the channel. They are
given in Table \ref{table3}. At low density $\Phi $ can be expanded in powers
of $k_F$
\begin{equation}
\Phi (v_F)=\Phi (k_F/\sqrt{M^2+k_F^2})\approx -{\frac{k_F^3}{3M^3}}+{\frac{%
3k_F^5}{10M^5}},  \label{apphi}
\end{equation}
which we will use explicitly in the following sections.

\section{Spectral density in the long-wavelength limit}
\label{sect:lwl}

The spectral density acquires contributions in the time-like region ($q>\nu $),
and also in the space-like region ($q<\nu $). The latter region is strictly
related to the presence of the
medium. Below we shall not be concerned with the modifications of the
time-like production cut by the medium. Since this cut is far away (in the
Fermi gas and for $q=0$ it starts at $\nu=2E_F$) its contribution to Eq. (\ref
{borel0}) is small. In addition, the limit of $q\rightarrow 0$ is regular in
the time-like region. This is not the case for the space-like region, where
special care is needed when $q\rightarrow 0$ \cite{BS}. The finite
contributions stemming from the space-like region are known as the
scattering terms, or Landau-damping terms. In this section we analyze in
detail the $q\rightarrow 0$ limit. Our results for scattering terms
for various channels are shown in Table \ref{table3}.

Since we have adopted the imaginary-time formalism in our approach, we have
been implicitly dealing with purely imaginary energies so far. Now, we are
going to do the analytic continuation to real energies. This can be done in
the standard way, by making the replacement $\nu \rightarrow \nu
_R+i\varepsilon $. In this way one obtains the retarded correlator for real
energies $\nu _R$. The imaginary part of the retarded correlator is
generated by the two logarithms appearing in Eq.~(\ref{xintegral}). The
imaginary part of the first logarithm is
\begin{eqnarray}
&&\ \pi \left[ \theta (-E_k+\sqrt{E_k^2+2kq+q^2}-\nu )-\theta (-E_k+\sqrt{%
E_k^2-2kq+q^2}-\nu )\right.  \nonumber \\
&&\ \left. +\theta (-E_k-\sqrt{E_k^2+2kq+q^2}-\nu )-\theta (-E_k-\sqrt{%
E_k^2-2kq+q^2}-\nu )\right] ,  \nonumber \\
&&  \label{th1}
\end{eqnarray}
where we have omitted the index ${}_R$ indicating that $\nu $ is real from now
on. For space-like momenta in the long-wavelength limit ($\nu <q\,\,%
\hbox{and}\,\,M>>\nu ,q\rightarrow 0$) only the first two terms contribute
and the imaginary part can be reduced to the expression
\begin{equation}
\pi \,\theta (k-|k_{-}(\nu ,q)|),\quad k_{-}(\nu ,q)=\frac \nu 2\sqrt{{\frac{%
4M^2}{q^2-\nu ^2}}+1}-\frac q2.\,
\end{equation}
In a similar way we deal with the second logarithm in (\ref{xintegral}) and
find that its imaginary part in the discussed region is
\begin{equation}
-\pi \,\theta (k-|k_{+}(\nu ,q)|),\quad k_{+}(\nu ,q)=\frac \nu 2\sqrt{{%
\frac{4M^2}{q^2-\nu ^2}}+1}+\frac q2.
\end{equation}
Consequently, the imaginary part of $\Pi ^{(i)}$ is equal to%
\footnote{%
It is still possible to change the variables in (\ref{pidef2}) in such a way
that the lower limit of integration in the two integrals is the same ($E_k=%
\half (qx-\nu )$ and $E_k=\half (qx+\nu )$ in the first and second integral,
respectively.) This leads to the compact integral representation of the
imaginary part valid for arbitrary temperature and chemical potential. For
vector correlators using the appropriate forms of ${\cal L}_V^{+}$ we
recover the result of Bochkarev and Shaposhnikov \cite{BS} in this way.}
\begin{equation}
\hspace{-1cm} \hbox{Im}\Pi ^{(i)}={\frac 1{128\pi q^3}}\left[
\,\,\int\limits_{|k_{-}|}^\infty {\frac{dk}{E_kk}}f(E_k,\mu ,\beta ){\cal L}%
_{(i)}^{+}-\int\limits_{|k_{+}|}^\infty {\frac{dk}{E_kk}}f(E_k,\mu ,\beta )%
{\cal L}_{(i)}^{-}\right] . 
\label{pidef2}
\end{equation}

For space-like momenta in the long-wavelength limit both $\nu $ and $q$
tend to zero but with their ratio is kept constant

\begin{equation}
q,\nu \rightarrow 0,\,\,\,\,\nu /q=\alpha \,\,\,\,\,\,(0\le \alpha <1).
\label{lwl}
\end{equation}
In this case we obtain 
\begin{equation}
k_{\mp }={\frac{\alpha M}{\sqrt{1-\alpha ^2}}}\mp {\frac q2}\equiv \kappa
\mp {\frac q2}  \label{k0}
\end{equation}
and the leading term of Eq. (\ref{pidef2}) is
\begin{eqnarray}
\!\!\!\!\!\!\!\!\!\!\!\!\hbox{Im}\Pi ^{(i)}(\alpha ) &=&\lim_{q\rightarrow 0}%
{\frac{1} {128\pi q^3}}\left\{ \int_\kappa^{k_F} {\frac{dk}{E_k k}}
f(E_k,\mu,\beta) \left[ {\cal L}_{(i)}^{+}(k,\alpha q,q)-{\cal L}%
_{(i)}^{-}(k,\alpha q,q)\right] \right.  \nonumber \\
\!\!\!\!\!\!\!\!\!\!\!\!&&+\left. {\frac q{2E_\kappa \kappa }}
f(E_{\kappa},\mu,\beta) \left[ {\cal L}_{(i)}^{+}(\kappa,\alpha q,q)+ {\cal L%
}_{(i)}^{-}(\kappa ,\alpha q,q)\right] \vphantom{ \int_{\kappa}^{k_F}}%
\right\} .  \label{pidef3}
\end{eqnarray}

Of special interest is the case of cold matter, where $\beta \rightarrow
\infty $. In this situation the distribution function (\ref{distr}) reduces
to the step function $\theta (\mu -E_k)=\theta (k_F-k)$, and formula (\ref
{pidef2}) becomes an integral over finite range of $k$ 
\begin{eqnarray}
\hbox{Im}\Pi ^{(i)}(\nu ,q)={\frac 1{128\pi q^3}} &&\left[ \theta
(k_F-|k_{-}|)\int_{|k_{-}|}^{k_F}{\frac{dk}{E_kk}}{\cal L}_{(i)}^{+}(k,\nu
,q)\right.  \nonumber \\
&&\left. -\theta (k_F-|k_{+}|)\int_{|k_{+}|}^{k_F}{\frac{dk}{E_kk}}{\cal L}%
_{(i)}^{-}(k,\nu ,q)\right] .  \label{pidef4}
\end{eqnarray}

In all the cases we consider ${\cal L}_{(i)}^{\pm }$ turns out to be a
simple polynomial and the integrals over $k$ in (\ref{pidef4}) can be easily
performed. We do not present the results of such integrations here, since
the final expressions are rather lengthy. Moreover, we are concentrated on
the long-wavelength limit, where further simplifications are possible. In
this case

\begin{equation}
\theta (k_F-|k_{\mp }|)=\theta (v_F-{\frac \nu q})=\theta (v_F-\alpha ),
\label{theta}
\end{equation}
and the zero-temperature limit is given by Eq. (\ref{pidef3}) with $%
f(E_k,\mu ,\beta )$ replaced by $\theta (v_F-\alpha )$.

We note that $\hbox{Im}\Pi ^{(i)}(\nu ,q)/(\nu ^2-q^2)$ becomes proportional
to the delta function, $\delta (\nu ^2)$, in the limit (\ref{lwl}), in
agreement with Eq. (\ref{delta}) \cite{BS}. 
This can be seen explicitly from the integral 
\begin{equation}
\hspace{-1.5cm}\lim_{q\rightarrow 0}\int d\nu ^2{\frac{\hbox{Im}\Pi
^{(i)}(\nu ,q)}{(\nu ^2-q^2)}}f(\nu ^2) =\int_0^1{\frac{2\alpha
\,d\alpha }{\alpha ^2-1}}\hbox{Im}\Pi ^{(i)}(\alpha )f(0)\equiv
R_{(i)} f(0),  
\label{delta1} 
\end{equation}
where $f(\nu ^2)$ is an arbitrary continuous function and $R_{(i)}$
describes the strength of the distribution. Our results for various
channels obtained for the zero-temperature case are shown in Table
\ref{table3}.

\section{Results for the vector channel}

In this section we discuss in some greater detail the case of the vector
channel. \label{sect:Vector}This channel is particularly important in view
of the experimental evidence of in-medium modification of light vector
mesons. First, we recall the well-known fact that on general grounds at $%
{\bf q}=0$ the transverse and longitudinal vector correlators are equal
\cite{chin}
\begin{equation}
\Pi _{V_T}(\nu ,0)=\Pi _{V_L}(\nu ,0).
\end{equation}
In order for this to hold, we note from Eq. (\ref{borel0}) that we must have

\begin{equation}
\lim_{q\rightarrow 0}\left( \Pi _{V_T}(q,q)-\Pi _{V_L}(q,q)\right) =\frac{%
R_{V_T}-R_{V_L}}\pi .  \label{check}
\end{equation}
Below we verify explicitly that indeed
this equality holds for the Fermi gas at
finite temperature.

The coefficients $A_{(i)},B_{(i)}$ and $C_{(i)}$ for the vector channel are
given in Table \ref{table2}. After inserting their explicit form into
Eq. (\ref{lpm}) we find
\begin{equation}
{\cal L}_{V_L}^{\pm }=-16k^2(\nu ^2-q^2)(4E_k^2+\nu ^2-q^2\pm 4E_k\nu )
\label{lpmvl}
\end{equation}
and
\begin{equation}
{\cal L}_{V_T}^{\pm }=-16k^2\left( 2k^2q^2-2E_k^2\nu ^2\mp 2E_k\nu (\nu
^2-q^2)-{\frac{\nu ^4}2}+{\frac{q^4}2}\right) .  \label{lpmvt}
\end{equation}
Substituting these equalities into Eq. (\ref{pidef3}) and (\ref{delta1})
yields, after elementary algebra, the following expression:
\begin{eqnarray}
\!\!\!\!\!\!\!\frac{R_{V_T}-R_{V_L}}\pi &=&\frac 3{\pi ^2}\int_0^1\alpha
^2d\alpha \int_\kappa ^{\infty}k\,{dk}\,f(E_k,\mu ,\beta )  \nonumber \\
\!\!\!\!\!\!\!\!\!\!\!\! &&+\frac{M^4}{\pi ^2}\int_0^1\frac{\alpha ^3d\alpha 
}{\left( \alpha ^2-1\right) ^2}{\frac 1{E_\kappa \kappa }}f(E_\kappa ,\mu
,\beta ),  \label{pidef6}
\end{eqnarray}
where $\kappa =\alpha M/\sqrt{1-\alpha ^2}$, and $E_\kappa =M/\sqrt{1-\alpha
^2}$. Next, we interchange the order of the $\alpha $ and $k$ integrations
in the first term in Eq. (\ref{pidef6}), and replace the integration
variable from $\alpha $ to $\kappa $ in the second term. As a result we
obtain exactly the integral for $\lim_{q\rightarrow 0}\Pi _{V_T}(q,q)$, as
given in Eq. (\ref{piqqvt}). Since $\Pi _{V_L}(q,q)=0$, it completes the
explicit check of Eq. (\ref{check}) in the Fermi gas.

Let us now turn to the discussion of the zero-temperature case. Through the
use of Eq. (\ref{lpmvl}) in (\ref{pidef4}) we recover Eq. (5.75) of Ref. 
\cite{chin}

\begin{equation}
\hbox{Im}\Pi ^{V_L}(\alpha )={\frac{\alpha (1-\alpha ^2)\theta (v_F-\alpha )%
}{2\pi }}E_F^2.  \label{pivl}
\end{equation}
The strength of the distribution function as defined in (\ref{delta}) or (%
\ref{delta1}) is

\begin{equation}
R_{V_L}=-{\frac{k_F^3}{3\pi E_F}}.  \label{svl}
\end{equation}
On the other hand, the explicit calculation in the transverse vector channel
gives

\begin{equation}
\hbox{Im}\Pi ^{V_T}(\alpha )={\frac{ \alpha  
(\alpha^2-v_F^2)\theta (v_F-\alpha )}{4\pi }} E_F^2
\label{pivt}
\end{equation}
and the corresponding strength is

\begin{equation}
R_{V_T}={\frac 1{6\pi }}\left[ {k_F^3 \over E_F}+3M^2\Phi (v_F)\right] .
\label{svt}
\end{equation}

The calculation of the imaginary part in other channels can be done in the
analogous way. In Table \ref{table3} we summarize the results for all the channels
giving the exact and approximate strength $R^{(i)}$, and the constant $%
W^{(i)}$ which determines the subtraction term $\Pi ^{(i)}(q,q)$.

To conclude this section, we comment on the use of the scattering term
in the existent calculations concerning vector mesons in medium. For
symmetric nuclear matter $k_F^3=3\pi ^2\rho /2$, where $\rho $ is the
baryon density.  Hence, in the limit (\ref{lwl}) one can write $-\half
\hbox{Im}\Pi _{V_L}/(\nu ^2-q^2)\rightarrow \rho _{sc}\delta (\nu
^2)/(8\pi )$, where $\rho _{sc}=2\pi ^2\rho /E_F$. The last result agrees 
with the normalization of the scattering term used in the literature
for {\em both} the longitudinal and transverse channels 
\cite{AHZ,HKL,Jin,Mallik,HL,HLS,LPM}. As we have shown, the scattering term 
for the transverse and longitudinal spectral densities is {\em different}. 
However, the presence of the subtraction constant $\lim_{q\rightarrow
0}\Pi_{V_T}(q,q)$ compensates this difference and, in effect, the
``cavalier'' approach of Ref. \cite{AHZ,HKL,Jin,Mallik,HL,HLS,LPM} indeed 
leads to correct results.

\section{Summary}

In this paper we have calculated the scattering term contributions, as well
as corresponding subtraction constants in various mesonic channels. We
stress that these subtraction constants are well-defined (unambiguous) in a
given many-body treatment. This is in contrast to the case of the vacuum,
where the subtraction constants, connected to renormalization, are not
well-defined. However, the presence of the medium does not bring new
divergences to the theory; in other words, medium effects are finite. This
is reflected by the finiteness of our constants $\Pi (q,q)$. We have also
shown the necessity of the subtraction constants for the consistency of the
approach: together with the scattering terms they lead to the required
equality of the correlators in the transverse and longitudinal channels when 
${\bf q\rightarrow 0}$.

As is well known, for finite ${\bf q}$ the longitudinal and transverse
correlators are no longer equal. Our expressions for the scattering term can
be straightforwardly extended to that case. The subtraction constants do not
depend on ${\bf q}$ in the Fermi-gas model.

The explicit calculations of this paper were done for the Fermi gas of
nucleons. In Table \ref{table3} we show our results, needed for the
construction QCD sum rules in these channels. We note, that the
calculation in the Fermi gas is a model calculation. Other effects,
such as meson production, rescattering, meson-exchange, {\em
etc. }will result in modifications of both the scattering terms and
the subtraction constants. These effects need not be small and should
be incorporated.
\begin{table}[tb]
\begin{tabular}{|p{1.5cm}|p{5cm}|p{3cm}|p{1.25cm}|}
\hline
\multicolumn{1}{|c|}{$i^{th}$ channel} & \multicolumn{1}{|c|}{$R^{(i)}$} & 
\multicolumn{1}{|c|}{$R^{(i)}$ at low density} & \multicolumn{1}{|c|}{$%
W^{(i)}$} \\ \hline\hline
$S$ & ${\frac{M^2}{\pi}}\Phi(v_F)$ & $-{\frac{k_F^3}{3\pi M}}$ & $-\half$ \\ 
\hline\hline
$P$ & 0 & 0 & $-\half$ \\ \hline\hline
$V_L$ & $-{\frac{k_F^3}{3\pi E_F}}$ & $-{\frac{k_F^3}{3\pi M}}$ & $0$ \\ \hline
$V_T$ & ${\frac{1}{6 \pi}}[{k_F^3 \over E_F} + 3 M^2 \Phi(v_F)]$ & ${%
\frac{k_F^5 }{15 \pi M^3}}$ & $\half$ \\ \hline\hline
$A_{qL}$ & $-{\frac{M^2}{\pi}} \Phi(v_F)$ & ${\frac{k_F^3}{3\pi M}}$ & $0$
\\ \hline
$A_L$ & $-{\frac{1}{3 \pi}}[{k_F^3 \over E_F} + 3 M^2 \Phi(v_F)]$ & $-{%
\frac{2 k_F^5 }{15 \pi M^3}}$ & $0$ \\ \hline
$A_T$ & ${\frac{ 1}{6 \pi}}[{k_F^3 \over E_F} - 3 M^2 \Phi(v_F)]$ & $%
{\frac{k_F^3}{3\pi M}}$ & $\half$ \\ \hline\hline
$T_{L^-}$ & ${\frac{2}{3 \pi}}[ {k_F^3 \over E_F}+ 3 M^2 \Phi(v_F)]$ & $%
{\frac{4 k_F^5 }{15 \pi M^3}}$ & $1$ \\ \hline
$T_{T^-}$ & $-{\frac{1}{3 \pi}}[{k_F^3 \over E_F}- 3 M^2 \Phi(v_F)]$ & $-{%
\frac{2 k_F^3}{3\pi M}}$ & $0$ \\ \hline
$T_{L^+}$ & $-{\frac{2 k_F^3 }{3 \pi E_F}}$ & $-{\frac{2 k_F^3}{3\pi M}}$ & $-1$
\\ \hline
$T_{T^+}$ & ${\frac{1}{3 \pi}}[{k_F^3\over E_F}+ 3 M^2 \Phi(v_F)]$ & $%
{\frac{2 k_F^5 }{15 \pi M^3}}$ & $0$ \\ \hline\hline
$VT_L$ & $0$ & $0$ & $0$ \\ \hline
$VT_T$ & $0$ & $0$ & $0$ \\ \hline
\end{tabular}
\caption{List of our results. The scattering terms are characterized by
the strength $R$ defined by Eq. (\ref{delta1}), whereas the subtraction
constants are characterized by the coefficient $W$ defined in Eq. (\ref{piqq}).
The function $\Phi$  is introduced in Eq. (\ref{phi}), and the limit of $R$ 
at low density is obtained from Eq. (\ref{apphi}).}
\label{table3}
\end{table}

\section{Appendix}

\label{sect:app1}Tensors $L^{\mu \nu }$, $T^{\mu \nu }$ and $Q^{\mu \nu }$
are defined as follows\footnote{%
Note that our definition of the $T$ and $L$ tensor has the sign chosen is
such a way that they are projection operators. Opposite convention is
frequently used, in order to make $\hbox{Im} \Pi_{V_L}/(\nu^2-q^2)$ and
$\hbox{Im} \Pi_{V_T}$ positive.} 
\begin{eqnarray}
T^{\mu \nu } &=&g^{\mu \nu }-u^\mu u^\nu -\frac{(q^\mu -q\cdot u\ u^\mu
)(q^\nu -q\cdot u\ u^\nu )}{q\cdot q-q\cdot u^2},  \label{tensors1T} \\
L^{\mu \nu } &=&-\frac{q^\mu q^\nu }{q\cdot q}+u^\mu u^\nu +\frac{(q^\mu
-q\cdot u\ u^\mu )(q^\nu -q\cdot u\ u^\nu )}{q\cdot q-q\cdot u^2},
\label{tensors1L} \\
Q^{\mu \nu } &=&\frac{q^\mu q^\nu }{q\cdot q}.  \label{tensors1Q}
\end{eqnarray}
The two projection operators $L^{\mu \nu }$ and $T^{\mu \nu }$ are both
transverse to $q$. In addition, $T^{\mu \nu }$ is transverse to $u^\mu $,
while $L^{\mu \nu }$ is longitudinal to $u^\mu $. The projection operator $%
Q^{\mu \nu }$ is longitudinal to $q$. It appears in the decomposition of the
axial vector correlator, since the axial-vector current is not conserved.
Tensors $L^{\mu \nu }$ and $T^{\mu \nu }$ can be obtained as a sum over
polarization vectors \cite{BFH}. Eqs. (\ref{tensors1T}-\ref{tensors1Q})
imply the following projection-operator relations: 
\begin{equation}
T_\mu ^{\,\,\alpha }T_\alpha ^{\,\,\nu }=T_\mu ^{\,\,\nu },\,\,\,L_\mu
^{\,\,\alpha }L_\alpha ^{\,\,\nu }=L_\mu ^{\,\,\nu },\,\,\,Q_\mu
^{\,\,\alpha }Q_\alpha ^{\,\,\nu }=Q_\mu ^{\,\,\nu },  \label{proj1}
\end{equation}
\begin{equation}
T_\mu ^{\,\,\alpha }L_\alpha ^{\,\,\nu }=T_\mu ^{\,\,\alpha }Q_\alpha
^{\,\,\nu }=Q_\mu ^{\,\,\alpha }L_\alpha ^{\,\,\nu }=0,  \label{perp1}
\end{equation}
\begin{equation}
T_\mu ^{\,\,\mu }=2,\,\,\,L_\mu ^{\,\,\mu }=1,\,\,\,Q_\mu ^{\,\,\mu }=1.
\label{norm1}
\end{equation}
The covariant decomposition of the vector and axial-vector correlators into
tensors (\ref{tensors1T}) -- (\ref{tensors1Q}) follows directly from the
Lorentz structure of expression (\ref{pidef}). In the rest-frame of the
medium, $u^\mu =(1,0,0,0)$, our definitions of the projection tensors reduce
to commonly used non-covariant expressions. 

We also give the appropriate projection tensors for other channels,
involving the tensor coupling. Such correlators have been considered in the
vacuum \cite{GRVW,BPKG}. Our expressions prepare grounds for the extension
of such calculations to finite density and temperature. The tensor
correlators can be decomposed in the basis of the projection tensors $T_{\mu
\nu ;\alpha \beta }^{(\pm )}$ and $L_{\mu \nu ;\alpha \beta }^{(\pm )}$,
where the sign $(\pm )$ refers to the parity of the excitation. The tensors $%
L_{\mu \nu ;\alpha \beta }^{(-)}$ and $L_{\mu \nu ;\alpha \beta }^{(+)}$
have the following structure 
\begin{eqnarray}
L_{\mu \nu ;\alpha \beta }^{(-)} &=&{\frac 1{2q^2}}\left[ L_{\mu \alpha
}q_\nu q_\beta +L_{\nu \beta }q_\mu q_\alpha -L_{\mu \beta }q_\nu q_\alpha
-L_{\nu \alpha }q_\mu q_\beta \right] ,  \label{tensors2min} \\
L_{\mu \nu ;\alpha \beta }^{(+)} &=&{\frac 1{2q^2}}\varepsilon _{\mu \nu
\sigma \tau }\varepsilon _{\alpha \beta \kappa \lambda }L^{\tau \kappa
}q^\sigma q^\lambda ,  \label{tensors2plus}
\end{eqnarray}
where $L_{\mu \nu }$ is defined in (\ref{tensors1L}). Tensors $T_{\mu \nu
;\alpha \beta }^{(-)}$ and $T_{\mu \nu ;\alpha \beta }^{(+)}$ follow from
equations (\ref{tensors2min}) and (\ref{tensors2plus}) with $L_{\mu \nu }$
replaced by $T_{\mu \nu }$. In analogy to Eqs. (\ref{proj1}-\ref{norm1}%
) we find 
\begin{equation}
T_{\mu \nu ;\alpha \beta }^{(\pm )}T_{(\pm )}^{\alpha \beta ;\sigma \tau
}=T_{\mu \nu ;}^{(\pm )\,\,\,\sigma \tau },\,\,\,\,L_{\mu \nu ;\alpha \beta
}^{(\pm )}L_{(\pm )}^{\alpha \beta ;\sigma \tau }=L_{\mu \nu ;}^{(\pm
)\,\,\,\sigma \tau },  \label{proj2}
\end{equation}
\begin{equation}
T_{\mu \nu ;\alpha \beta }^{(\pm )}L_{(\pm )}^{\alpha \beta ;\sigma \tau }=0,
\label{perp2}
\end{equation}
\begin{equation}
T_{\mu \nu ;}^{(\pm )\,\,\,\mu \nu }=2,\,\,\,\,L_{\mu \nu ;}^{(\pm
)\,\,\,\mu \nu }=1.  \label{norm2}
\end{equation}
We note that all products of the tensors with opposite parity vanish.

Tensors $L_{\mu ;\,\alpha \beta }$ and $T_{\mu ;\,\alpha \beta }$ appear in
the decomposition of the correlators in the vector-tensor channel. They are
defined as

\begin{equation}
L_{\mu ;\,\alpha \beta }=\sqrt{{\frac 1{2q^2}}}\left( L_{\mu \alpha }q_\beta
-L_{\mu \beta }q_\alpha \right) ,  \label{l3}
\end{equation}
and 
\begin{equation}
T_{\mu ;\,\alpha \beta }=\sqrt{{\frac 1{2q^2}}}\left( T_{\mu \alpha }q_\beta
-T_{\mu \beta }q_\alpha \right) .  \label{t3}
\end{equation}
The normalization of $L_{\mu ;\,\alpha \beta }$ and $T_{\mu ;\,\alpha \beta }
$ is induced by the requirement that the algebra of projection tensors is
closed. For example, the products of $L_{\nu ;\,\alpha \beta }$ with $L_\mu
^{\,\nu }$ and $L_{\,\,\,\,\, \alpha \beta }^{\sigma \tau ;}$ are

\begin{equation}
L_{\mu ;\,\alpha \beta } L_{\,\,\,\quad \sigma \tau }^{\alpha
\beta ;}= L_{\mu ;\,\sigma \tau ,\quad }\ L_\mu ^{\, \,\nu } L_{\nu
;\,\alpha \beta }= L_{\mu ;\,\alpha \beta }.
\end{equation}

\end{document}